\documentclass[a4paper]{jpconf}
\usepackage{graphicx}

\begin{document}
\title{Particle Showers in a Highly Granular Hadron Calorimeter}

\author{Frank Simon, for the CALICE Collaboration}

\address{Max-Planck-Institut f\"ur Physik and Excellence Cluster `Universe', Munich, Germany}

\ead{frank.simon@universe-cluster.de}

\begin{abstract}
The CALICE collaboration has constructed highly granular electromagnetic and hadronic calorimeter prototypes to evaluate technologies for the use in detector systems at a future Linear Collider. The hadron calorimeter uses small scintillator cells individually read out with silicon photomultipliers. The system with 7608 channels has been successfully operated in beam tests at DESY, CERN and Fermilab since 2006, and represents the first large scale tests of these devices in high energy physics experiments. The unprecedented granularity of the detector provides detailed information of the properties of hadronic showers, which helps to constrain hadronic shower models through comparisons with model calculations. We will discuss results on longitudinal and lateral shower profiles compared to a variety of different shower models, and present studies of the energy reconstruction of hadronic showers using software compensation techniques. 
\end{abstract}

\section{The CALICE Calorimeters}

The goal of the CALICE experimental program is to establish novel technologies for calorimetry in detectors at a future linear $e^+e^-$ collider, and to record electromagnetic and hadronic shower data with unprecedented three dimensional spatial resolution for the
validation of simulation codes and for the test and development of reconstruction algorithms. Such highly granular calorimeters are necessary to achieve an unprecedented jet energy resolution at the International Linear Collider \cite{:2007sg} using particle flow algorithms \cite{Brient:2002gh, Morgunov:2002pe, Thomson:2009rp}.

\begin{figure}
\centering
\includegraphics[width=0.95\textwidth]{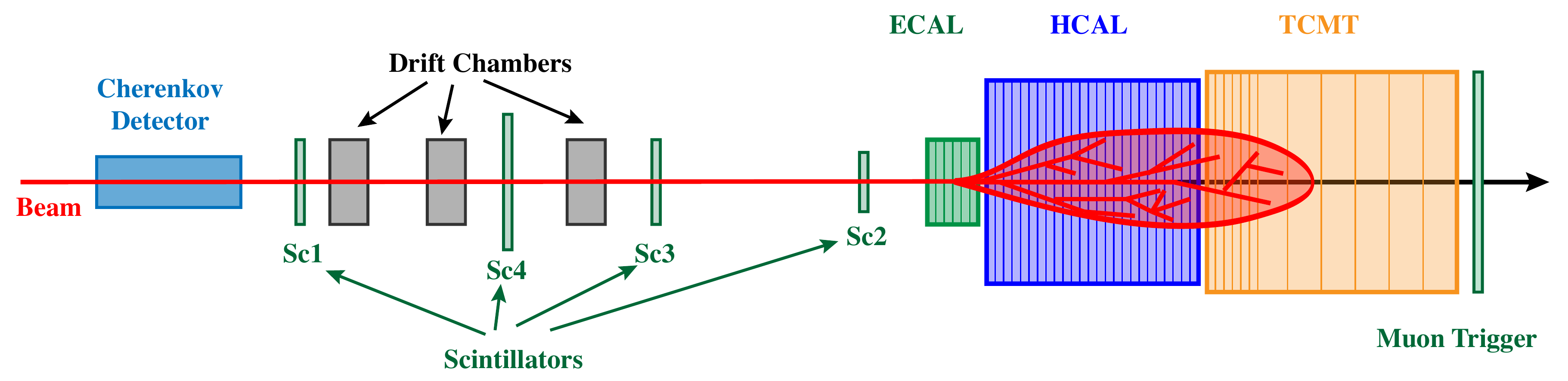}
\caption{Schematic of the CALICE experimental setup at
CERN, with electromagnetic and hadronic calorimetry as
well as a tail catcher and muon tracker downstream of the
calorimeters.}
\label{fig:CALICESetup}
\end{figure}

The CALICE test beam setup consists of a silicon-tungsten electromagnetic calorimeter (ECAL), an analog scintillator-steel hadron calorimeter (AHCAL) and tail catcher/muon tracker (TCMT), the latter two both with individual cell readout by silicon photomultipliers (SiPMs) \cite{Bondarenko:2000in}. This setup has been tested extensively in electron, muon, and hadron beams at CERN and at the Meson Test Beam Facility at Fermilab. Figure \ref{fig:CALICESetup}  shows the schematic setup of the CALICE detectors in the CERN H6 test beam area, where data was taken in 2006 and 2007. The currently ongoing and the future program include the study of alternative technologies, such as an ECAL using scintillator strips with SiPM readout, digital hadron calorimeters with active layers based on gas detectors, and the investigation of alternative absorber materials for the hadron calorimeter for multi-TeV colliders.

The performance of the current setup for hadronic showers is driven by the analog HCAL \cite{Adloff:2010hb}, which is  a 38 layer sampling calorimeter with 5 mm thick scintillator layers sandwiched by 2 cm of steel. The lateral dimensions are roughly 1$\times$1 m$^2$, the total thickness amounts to 5.3 nuclear interaction lengths. The active layers are built out of scintillator tiles with sizes ranging from 30$\times$30 mm$^2$ in the core of the detector to 120$\times$120 mm$^2$. Each tile is read out by a built-in SiPM, produced by the MEPhI/PULSAR group \cite{Bondarenko:2000in}. In total, the calorimeter has 7608 channels. This high granularity allows for detailed studies of hadronic showers, and lends itself to energy reconstruction algorithms using information of the shower substructure.

In the present paper, hadron data taken at CERN in 2007 with the complete installation of the silicon tungsten ECAL, the analog HCAL and the TCMT, are being discussed, focusing on results from the hadron calorimeter. 

These experimental results are confronted with detailed simulations performed within the {\sc Geant4} framework. For the simulations, a variety of different hadronic shower models, implemented as different ``physics lists''  \cite{Geant4Physics} within {\sc Geant4}, are used.  The simulation of the interaction and propagation of particles is coupled with a detailed simulation of the detector response, including the effects of photon statistics, saturation effects in the SiPM and in the scintillator, and the integration time of the electronics. The comparison of simulation and measurement will provide insight for the further development of {\sc Geant4} physics lists.

\section{Shower Profiles}

\begin{figure}
\centerline{\includegraphics[width=0.5\textwidth]{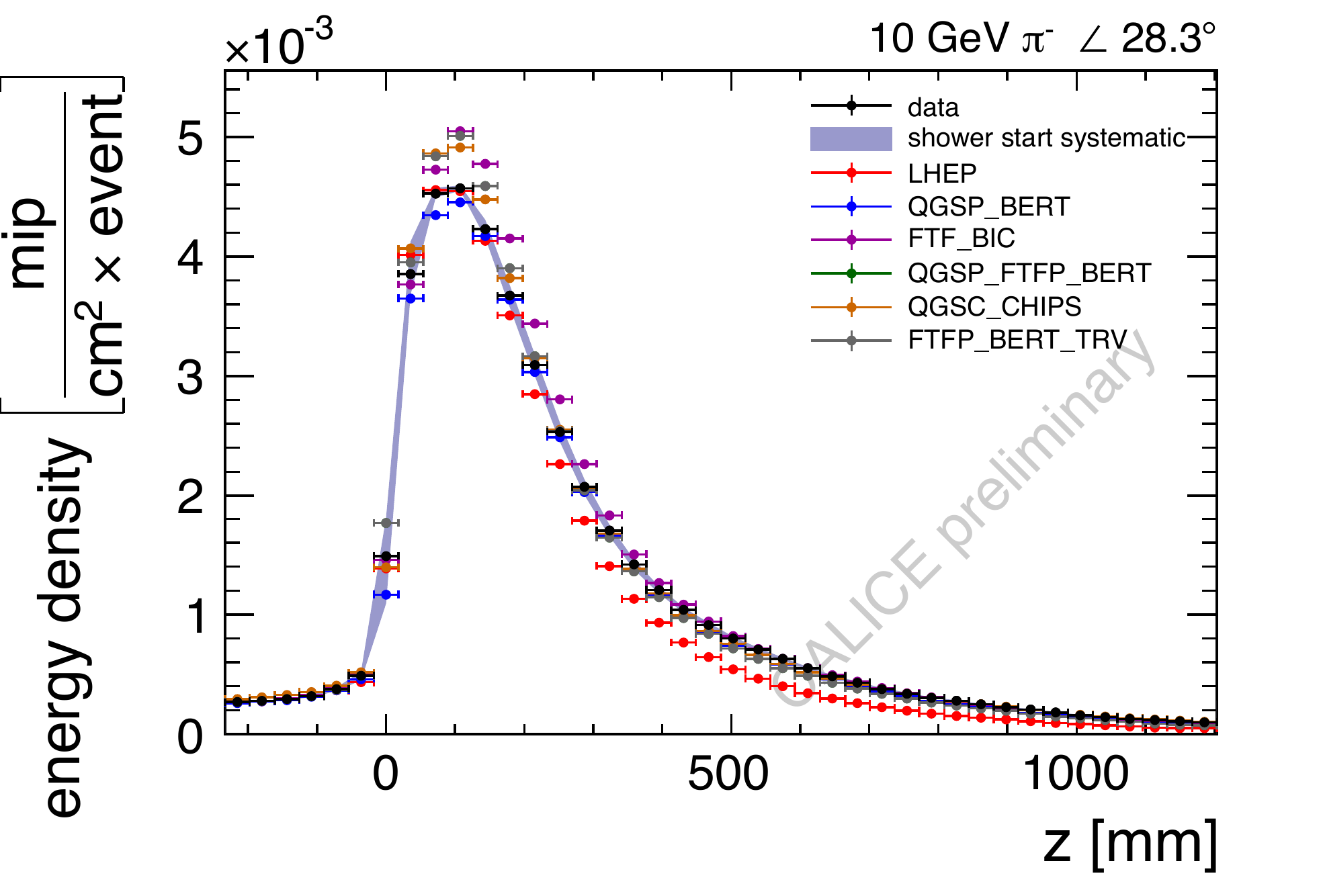} \includegraphics[width=0.5\textwidth]{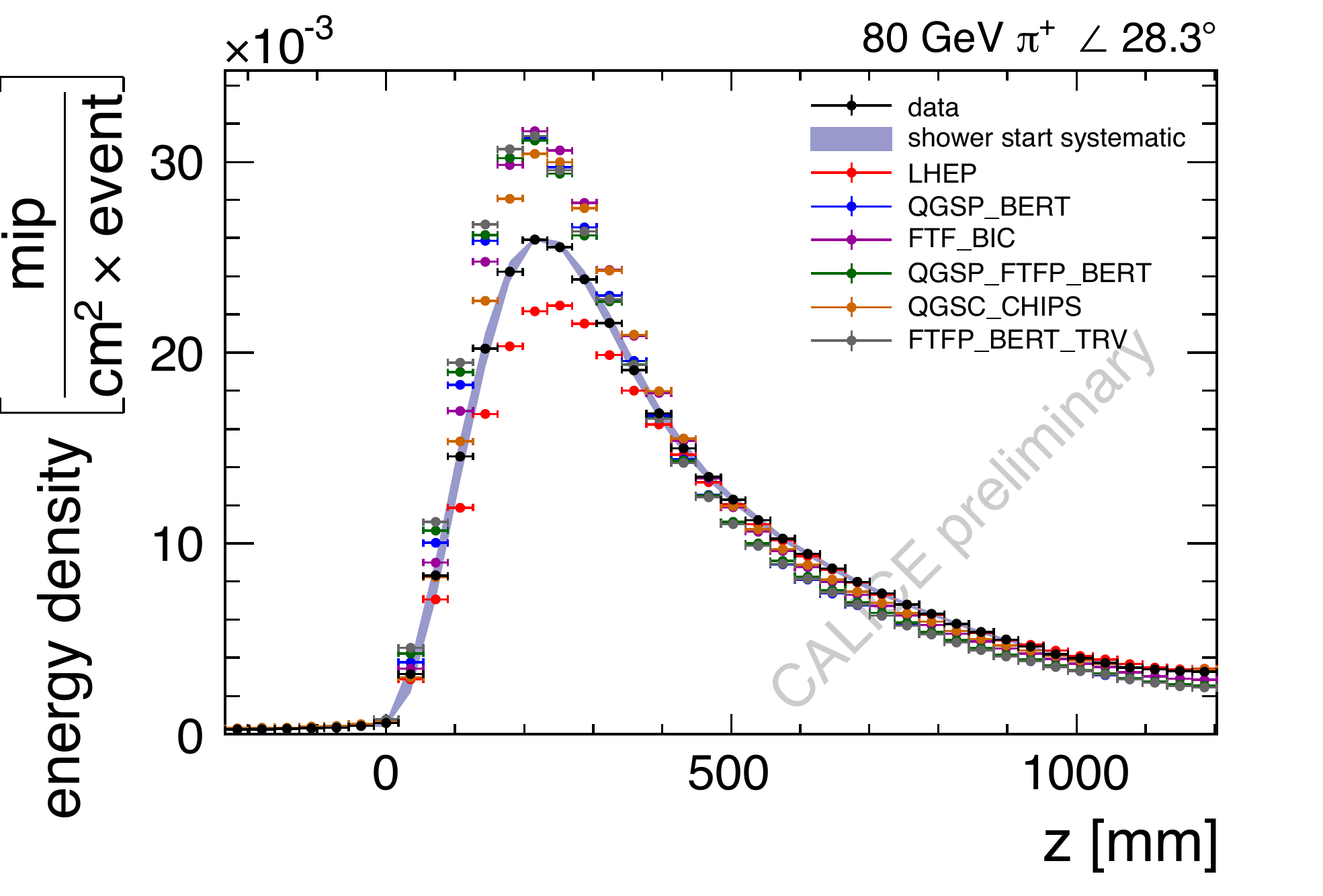}}
\caption{Longitudinal profile of hadronic showers of 10 GeV $\pi^-$ (left) and 80 GeV $\pi^+$ (right) compared to various hadronic shower models.}\label{fig:LongProfile}
\end{figure}

The high granularity of the calorimeter, with layer-wise longitudinal and fine lateral segmentation, allows precise measurements of the topology of hadronic showers. Particularly powerful is the possibility to identify the starting point of the shower, by measuring the position of the first inelastic reaction of the incoming beam particle via a local increase in energy deposit and hit multiplicity over that expected from the passage of a minimum ionizing particle.  This allows the measurement of shower properties with respect to the starting point, eliminating the smearing introduced by the exponential distribution of the point of the first interaction, resulting in a significant increase in sensitivity to the longitudinal evolution of the shower.

Figure \ref{fig:LongProfile} shows the longitudinal shower profile measured for 10 GeV negative and for 80 GeV positive pions relative to the shower start., measured with the detector inclined by approximately 30$^\circ$ with respect to the beam axis. The measurement is compared to a variety of different hadronic shower models. For the 10 GeV case, in particular the theory driven model \texttt{QGSP\_BERT}, favored by the LHC calorimeter test beams, provides a satisfactory description of the measurement, while other models show discrepancies in the region of the shower maximum and / or in the tail of the profile. At the higher energy point, all models show discrepancies with the data in particular in the shower maximum region.

Such measurements have the potential to elucidate the longitudinal dependence of the relative contributions of different components of hadronic showers, such as nuclear fragments,  electromagnetic subshowers  and hadrons, through detailed comparisons to simulations, as demonstrated already for the CALICE ECAL \cite{Adloff:2010xj}. Similar studies for the analog HCAL are currently in preparation.

\begin{figure}
\centerline{\includegraphics[width=0.5\textwidth]{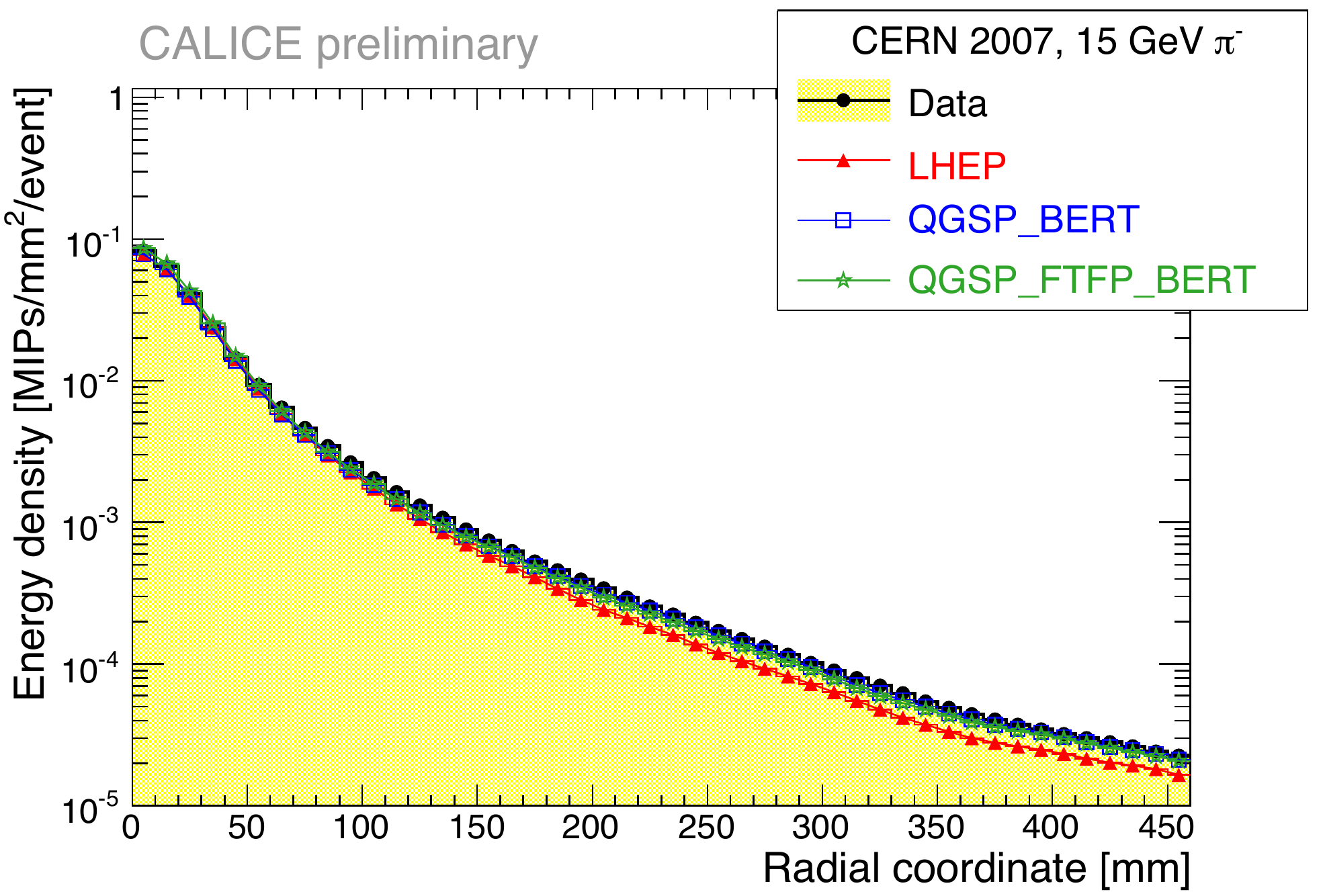} \hfill \includegraphics[width=0.5\textwidth]{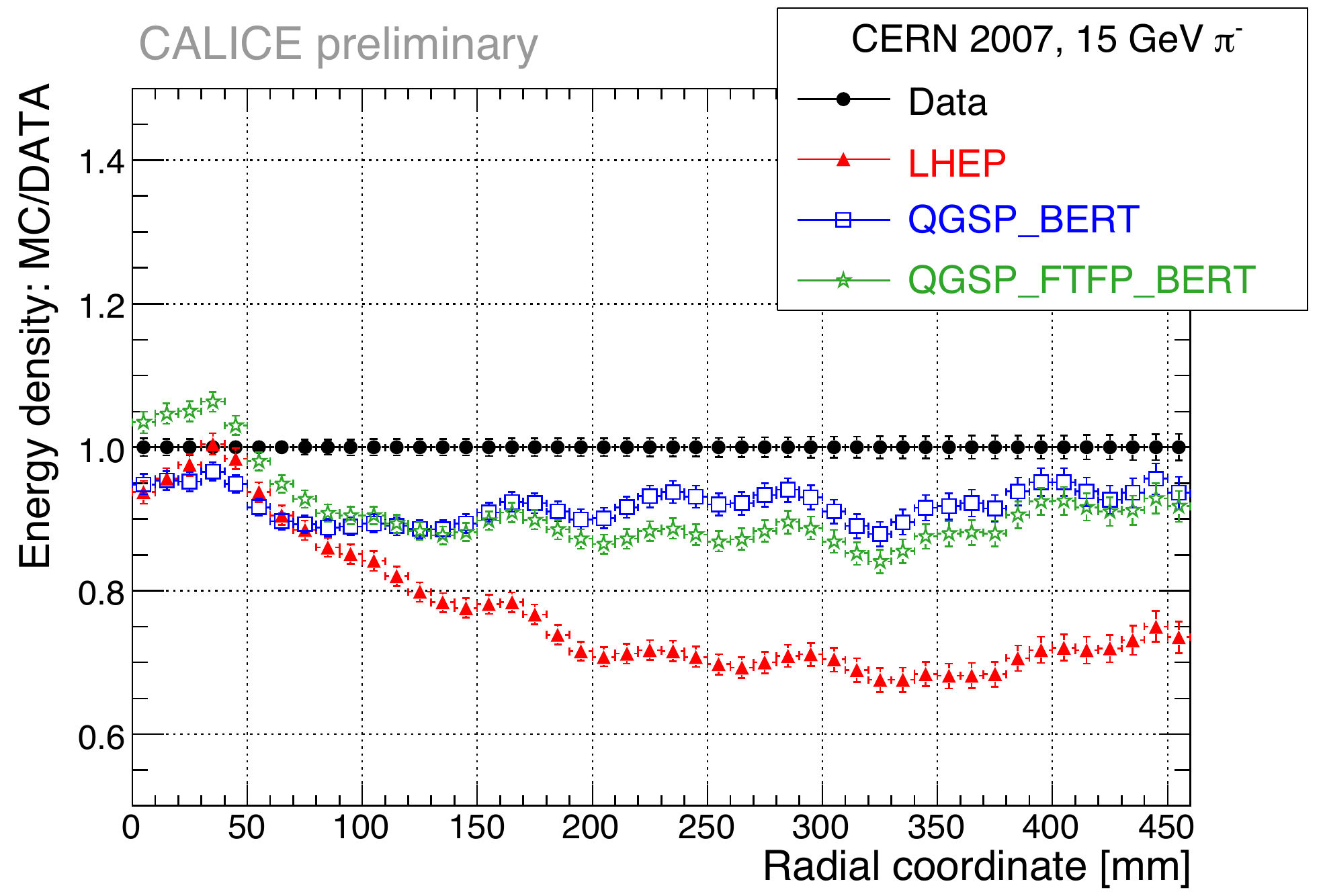}}
\caption{Lateral profile of showers of 15 GeV $\pi^-$. {\it Right:} Energy density as a function of radial coordinate. {\it Left:} Ratio of simulations based on different physics lists and data.}
\label{fig:LatProfile}
\end{figure}

The performance of Particle Flow Algorithms is strongly influenced by the capability for the separation of particle showers in the calorimeter. For the evaluation of such algorithms with simulations, a realistic description of the lateral shower size is thus of importance. Figure \ref{fig:LatProfile} ({\it left}) shows the transverse shower profile, expressed by the energy density as a function of the radial coordinate. The radial distance of energy deposits to the particle axis is determined using the particle tracking information from the drift chambers in the CALICE beam setup, allowing sub-cellsize resolution. 

As for the longitudinal profiles, the data are confronted with different hadronic shower models, with three such models shown in the figure. Figure \ref{fig:LatProfile} ({\it right}) shows the ratio of the simulation predictions and the measurement, showing discrepancies for all considered models. The models shown here describe the shower core within approximately 5\%, with larger deviations observed in the outer regions, in particular in the case of {\tt LHEP}.

\section{Shower Substructure}

The extremely high granularity of the CALICE calorimeters also enables measurements of the substructure of showers well beyond averaged shower profiles discussed above. Such studies test the quality of shower simulations in finer detail, by requiring a realistic modeling of the shower evolution on the particle level, not only in terms of the averaged overall energy distribution.  

One such possibility is the detection of individual track segments within the shower, caused by single charged particles. Since such tracks behave as minimum-ionizing particles, they are a valuable calibration tool for a future linear collider calorimeter, enabling the monitoring of the calibration with regular physics data \cite{Lu:2009ze}. In the CALICE analog HCAL, such minimum-ionizing track segments are found from isolated hits, local energy deposits that do not have energy in next neighboring cells in the same calorimeter layer \cite{Weuste:2010sm}. This ensures a high purity of single particle tracks, but limits the track finding to the outer, less dense regions of the hadronic showers. The minimum length of tracks was 6 layers, corresponding to 0.85 $\lambda_I$, also imposed to increase the purity of the sample.

\begin{figure}
\centerline{\includegraphics[width=0.48\textwidth]{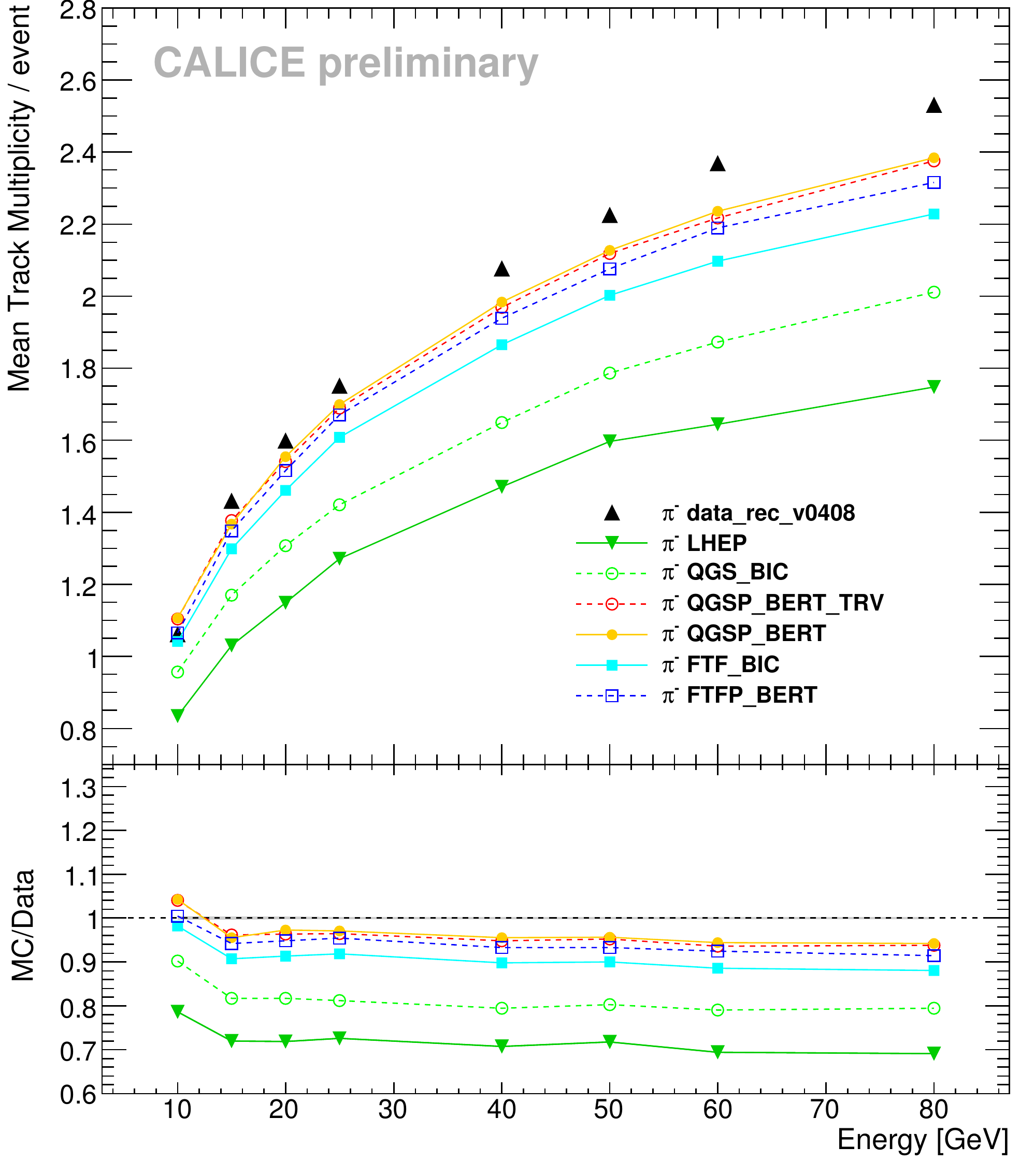} \hfill \includegraphics[width=0.48\textwidth]{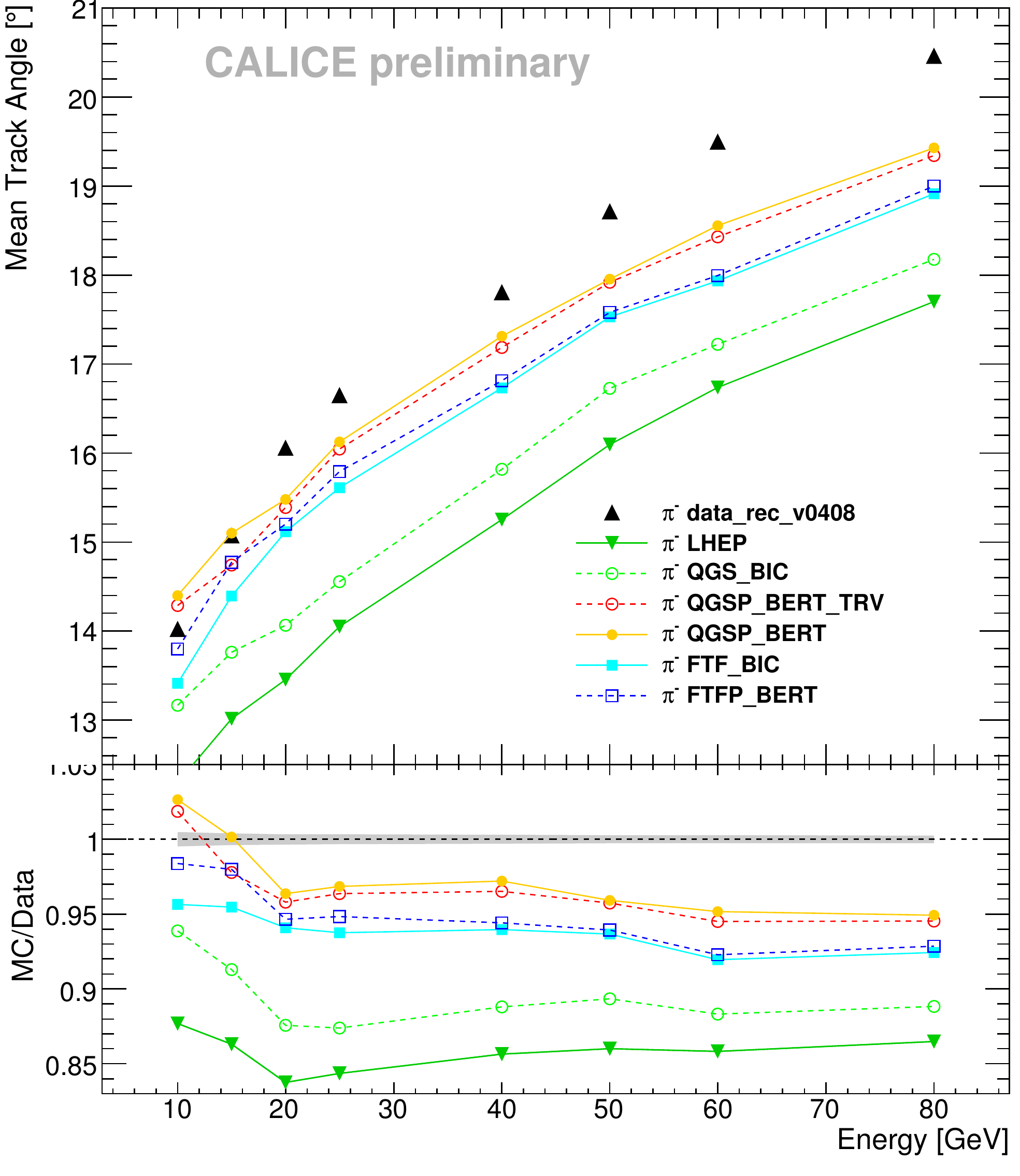}}
\caption{Average track multiplicity (left) and mean track angle (right) as a function of beam energy. Also shown are simulations based on different physics lists, and the ratio of simulations to data.}\label{fig:Tracks}
\end{figure}

Figure \ref{fig:Tracks} shows the average multiplicity and the mean angle of the found tracks for negative pions over a wide energy range. All considered shower models reproduce the general trend of increasing track multiplicity and increasing track angle with energy. While {\tt QGSP\_BERT} and the {\tt FTF} - based models reproduce the observations typically within 10\% and 5\% for the track multiplicity and angle, respectively, {\tt LHEP} and {\tt QGS\_BIC} show larger discrepancies. These latter two models thus produce a too small particle multiplicity in the outer parts of the showers, and include not enough particle production at large angles. {\tt QGSP\_BERT}, the physics list favored by the LHC calorimeter test beams, provides the best overall description of the track segments.

\section{Energy Reconstruction and Software Compensation}

The detailed information of hadronic showers provided by imaging calorimeters is also beneficial for the energy reconstruction. The resolution of non-compensating hadronic calorimeters, such as the CALICE calorimeters, suffers from the fact that such detectors respond differently to electromagnetic and to purely hadronic showers. Coupled with the large event-by-event fluctuations of the electromagnetic component and with the energy dependence of the mean electromagnetic fraction, this leads to a limited energy resolution and a non-linear response of such calorimeters.

The high granularity of the CALICE calorimeters provides excellent conditions for software compensation techniques. Such energy reconstruction algorithms treat electromagnetic and hadronic energy deposits separately to improve the energy resolution and to recover the linearity of the response. Since typical non-compensating calorimeters show a higher response to electromagnetic than to hadronic energy deposits ($e/h > 1$), electromagnetic deposits are assigned a lower weight in the overall energy sum of the shower. 

For the analog HCAL, two approaches to this technique have been studied: A local and a global software compensation algorithm. Both use the fact that electromagnetic energy deposits tend to have a higher energy density than purely hadronic ones. The local software compensation technique \cite{Simon:2009bt} uses the energy of each cell as a measure of the local energy density at that point in the shower, and chooses a weight for the cell in question according to this. The global technique \cite{Seidel:2010sf} uses the overall energy density of the shower, calculated from the shower volume found by a clustering algorithm to select one global weight for the shower, essentially corresponding to a determination of the overall electromagnetic content of the shower. These weights are applied relative to the naive energy calculation, which is a simple sum of the amplitudes of all cells above the noise threshold, multiplied by a constant factor to transform from the calibration scale in units of minimum ionizing particles (MIP) to the GeV scale. In both software compensation techniques considered here, the final algorithm does not require knowledge of the true beam energy, making it applicable over a wide energy range.

 \begin{figure}
\centerline{\includegraphics[width=0.5\textwidth]{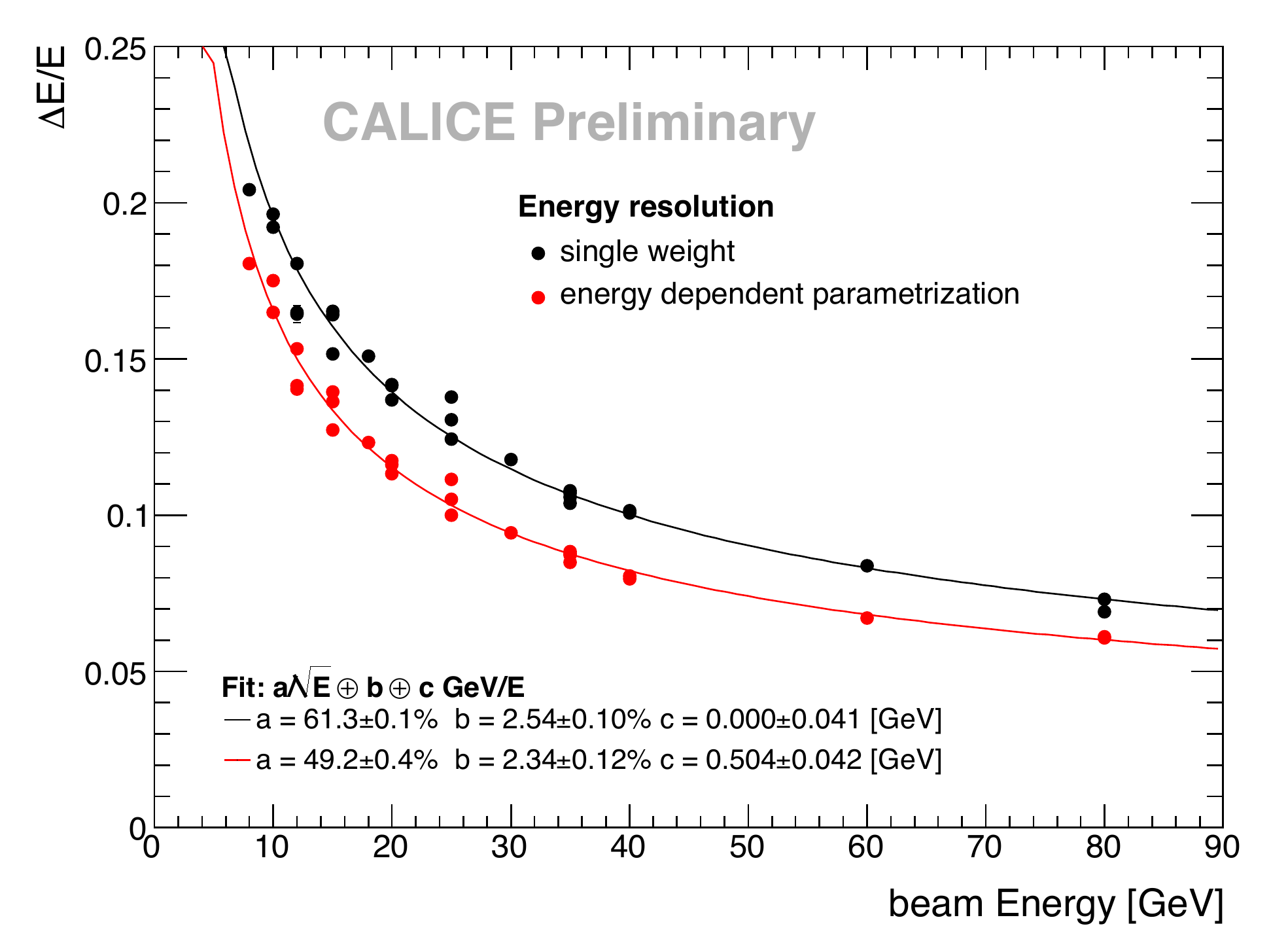} \hfill \includegraphics[width=0.5\textwidth]{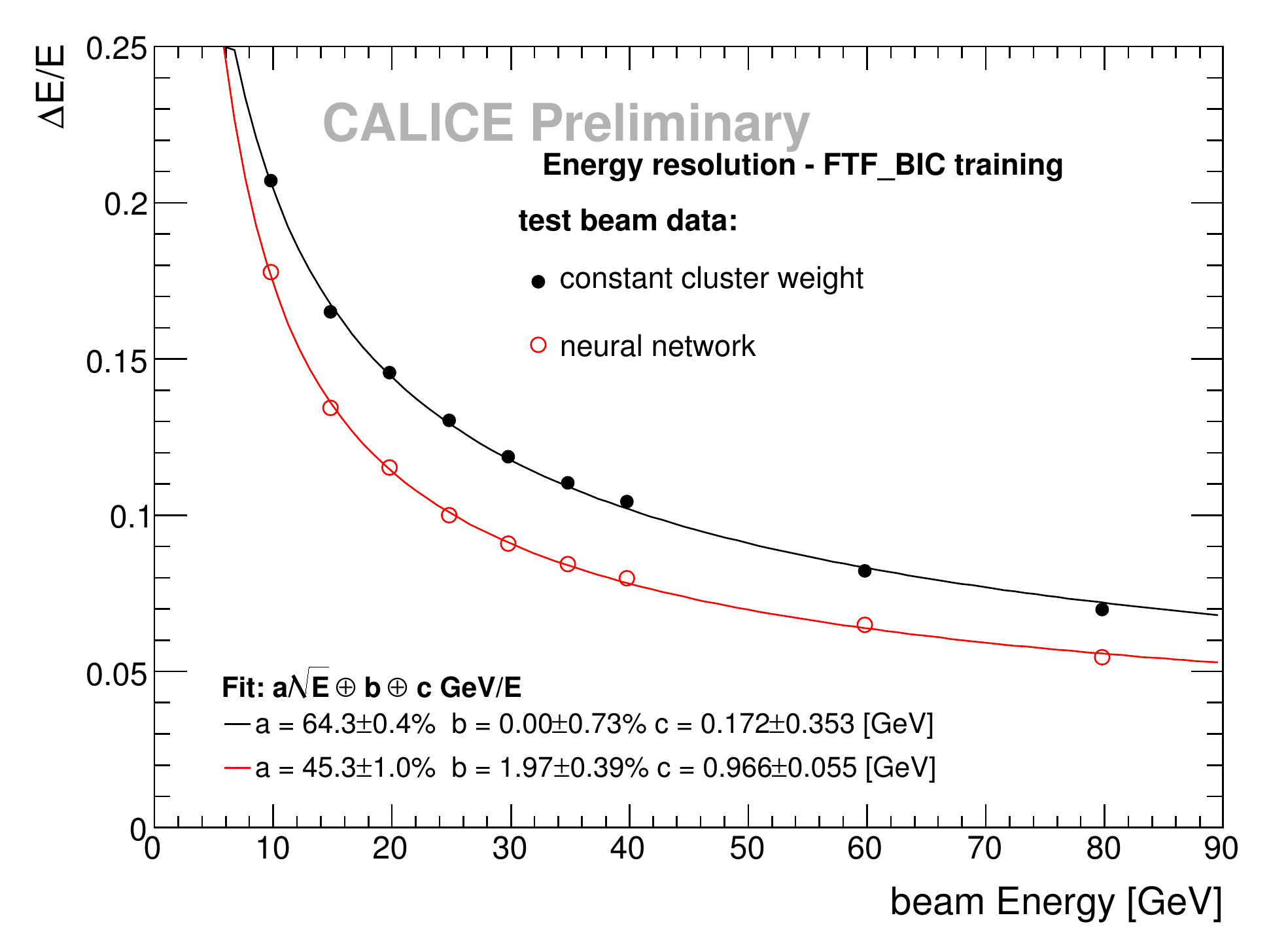}}
\caption{Energy resolution for hadrons with and without software compensation. {\it Left:} Local software compensation, with weights determined from data, applied to the full CALICE setup of ECAL, HCAL and TCMT. {\it Right:} Global software compensation on clustered data, with a neural network trained on simulations with the {\tt FTF\_BIC} physics list, applied to showers in the HCAL and TCMT.}\label{fig:Resolution}
\end{figure}

For a first study of local software compensation, the single-cell energy density based weights were determined from an independent real data set using a $\chi^2$ minimization technique \cite{Simon:2009bt}. Due to energy-dependent factors of the shower structure, such as changing mean electromagnetic fractions and increasing core energy density for increasing particle energy, these weights are energy dependent. A parametrization of this energy dependence was derived from the first set of weights obtained from the minimization technique, making the software compensation algorithm independent from the beam energy. This algorithm was then applied to the complete CALICE setup, consisting of the ECAL, HCAL and TCMT, without any requirements on shower containment. Figure \ref{fig:Resolution}{\it(left)} shows the energy resolution obtained with the local algorithm, compared to the reconstruction with one constant conversion factor. The compensation algorithm improves the resolution by up to 20\%, yielding a stochastic term of approximately 50\%/$\sqrt{E}$. 

The global software compensation was studied for showers in the HCAL and TCMT, by requiring an energy deposit in the ECAL consistent with a single minimum-ionizing particle. No further requirements on shower containment were made. With a simple clustering algorithm, the cells belonging to a particle shower were identified, and from this shower parameters such as the total energy and the shower volume were extracted. These parameters were used as inputs for a neural network to obtain an optimized estimate of the true particle energy. For this estimation, the energy density, given by the total shower energy divided by the shower volume, is of particular importance. The neural network was trained with a simulated data set with a quasi-continuous energy distribution to avoid artifacts from the discrete particle energies used in the test beam.  Several different physics lists were tested, with {\tt FTF\_BIC} giving the best results. Figure \ref{fig:Resolution}{\it(right)} shows the energy resolution obtained with the neural network, compared to the resolution for clustered showers with one constant conversion factor. The neural network achieves an improvement of up to 25\%. The fact that the energy resolution without software compensation in the global method is worse than in the case of the local method is due to additional effects from the clustering applied here. 

\section{Summary}

The CALICE collaboration has acquired sizeable data sets of hadronic showers with a highly granular hadronic calorimeter using a large sample almost 8\,000 scintillator tiles read out with silicon photomultipliers, operated stably over several data taking periods. These data provide detailed information on the overall shape and on the substructure of hadronic showers. The comparison of longitudinal and transverse shower profiles to simulations with a variety of different hadronic shower models provides input for a further development of these models. A first study of the shower substructure using identified minimum ionizing track segments has shown qualitative agreement for some of the {\sc Geant4} physics lists under consideration, underlining the level of detail already reached in these models. The high granularity of the detectors also proves to be a powerful tool for the improvement of the energy resolution with software compensation algorithms, both on a local cell-by-cell and on a global shower level. The remarkable improvement in performance with a neural network trained with simulations suggests that the global features of hadronic showers in the CALICE calorimeters, such as the overall energy density and its evolution with particle energy, are already well reproduced by the used physics list within {\sc Geant4}.

\section*{References}
\bibliography{CALICE}

\end{document}